# Phase retrieval technique from regions of unresolvable fringe density using Phase Shifting Interferometry and Liquid Crystal on Silicon (LCOS) Spatial Light Modulator (SLM)

Indrani Bhattacharya[1,2]
[1]Institute of Production Metrology (IPROM), Technische Universität, Braunschweig, Germany
[2]Department of Applied Optics and Photonics, University of Calcutta, Kolkata, India
bindrani03@gmail.com

**Abstract**
This paper presents a simple method to recover phase information from regions of unresolvable fringe density, specifically, at the camera corners, where interference fringes are not resolved by the camera using Phase Shifting Interferometry and Liquid Crystal on Silicon (LCOS) Spatial Light Modulator (SLM). The method works well with wavefront compensation. This work is a continuation and advancement of reported research work [14].

## 1. Introduction

Retrieving the phase information of a surface from an interferogram is a fundamental task in optical metrology, with applications in surface profiling, wavefront sensing, and quality control. The 4-step Phase Shifting Algorithm (PSA) is a well-established method for this purpose, offering robustness and accuracy by capturing successive interferograms with known phase shifts and extracting the phase via trigonometric computation [1][2]. However, this method's reliability diminishes in areas with high fringe density—particularly at the corners or near sharp gradients—where fringes become unresolved due to aliasing, low contrast, or excessive modulation frequency. These conditions result in degraded phase information, manifesting as noise, artifacts, or outright failure of the retrieval process [3][4].

Fringe resolution failure in such regions stems from several well-documented causes. Low fringe visibility—often due to weak illumination, out-of-focus regions, or spatially uniform lighting—can flatten the interferogram, making fringes indiscernible. Overexposure or saturation causes bright areas to clip to white, thereby erasing fringe modulation. In low signal-to-noise ratio (SNR) environments, random noise or speckle patterns dominate, obscuring the sinusoidal fringe structure required for phase computation. Steep phase discontinuities and wrapping artifacts introduce instability in phase unwrapping algorithms, particularly where sharp jumps exceed the unwrapping threshold [6][7]. In these cases, the sinusoidal variation assumed by PSA breaks down—leading to unreliable atan2 computations and failure of phase unwrapping routines [8]. Consequently, a careful analysis of fringe visibility and signal quality is essential for ensuring robust and accurate phase retrieval in practical applications [9].

This paper is a continuation and advancement of our recently reported research work [14], where we reported a simple and novel technique for interferometric surface measurement using a Liquid Crystal on Silicon (LCOS) Spatial Light Modulator (SLM) as phase shifter and wavefront-compensator simultaneously. This paper deals with a phase retrieval technique from regions of unresolvable fringe areas, especially in the corners, where the fringe density is too high and the spatial resolution of the camera is not enough to sample them correctly, leading to aliasing where the fringe pattern is under-sampled and low visibility where the fringes blend and appear as flat regions.

## 2. Theoretical Background

In the 4-step PSA, four intensity images are recorded with phase shifts (typically of $\pi/2$) introduced between them. The recorded interferogram intensity intensity $I_k(x,y)$ at each pixel $(x,y)$ is:

$$I_k(x,y) = I_0(x,y) + I_m(x,y)\cos[\phi(x,y) + \delta_k] \qquad (1)$$

where $I_k(x,y)$ is the is the observed intensity at pixel (x, y), $I_0(x,y)$ is the background or average intensity, $I_m(x,y)$ is the fringe modulation, $\phi(x,y)$ is the phase to retrieve, $\delta_k$ is the global phase shift for the $k$ th interferogram, i.e., 0, $\pi/2$, $\pi$ and $3\pi/2$ in our case. Using the four interferograms recorded with uniform phase shifts of $\delta=90^0=\pi/2$, we can retrieve $\phi(x,y)$ using the formula mentioned in [14],

$$\phi(x,y) = \tan^{-1}\left[\frac{I_4 - I_2}{I_1 - I_3}\right] \qquad (2)$$

where $I_1$, $I_2$, $I_3$ and $I_4$ are the interferograms recorded with uniform phase shifts of 0, $\pi/2$, $\pi$ and $3\pi/2$ respectively.

The Modulation amplitude $B(x,y)$ is computed to assess reliability of the fringe using the formula:

$$B(x,y) = \frac{1}{2}\sqrt{(I_1 - I_3)^2 - (I_2 - I_4)^2} \qquad (3)$$

When the spatial fringes are unresolved, the temporal intensity variation at each pixel due to phase shifting can be used ensuring that the modulation depth, $B(x,y)$ is non-zero. Modulation amplitude computed from the phase-shifted images is used a key indicator of signal reliability and a normalized value of $B(x, y)$ is used to create a reliability mask. Combination of LCOS and modulation amplitude masks are used to isolate the trustworthy phase data. In the next step, the 4-step Phase Shifting Algorithm (PSA) is performed to compute wrapped phase within the range $[0, 2\pi]$ and for unwrapping, an algorithm by Herraez et al. [10] in a Matlab implementation freely provided by Kasim [11] is used.

## 3. Experimental Arrangement

The schematic diagram of the experimental arrangement of LCOS SLM Phase Shifting Interferometer with Twyman-Green configuration is shown in **Figure 1.**

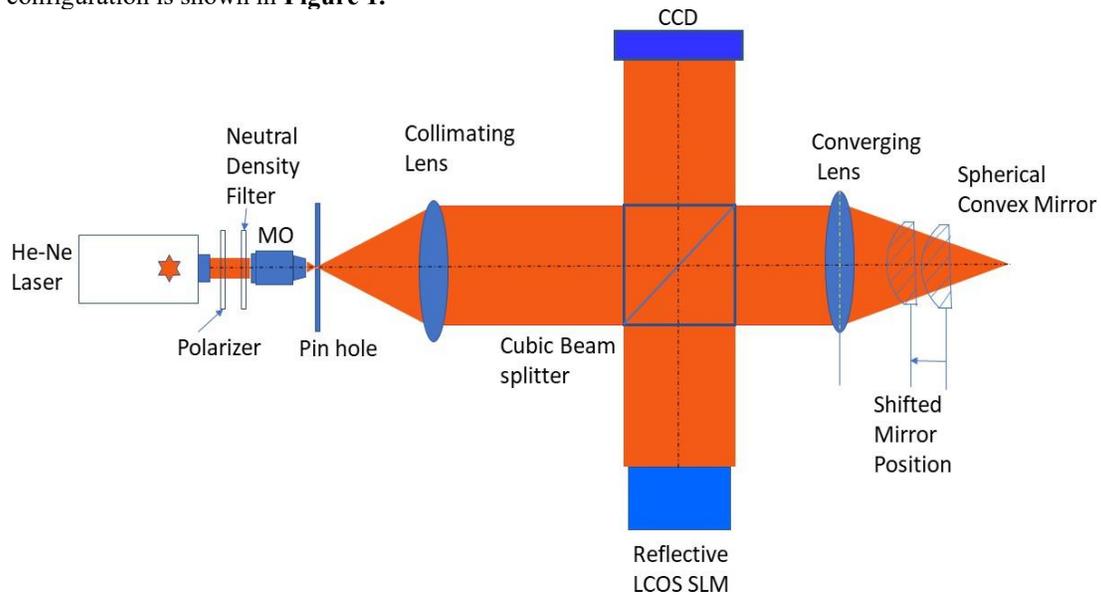

**Figure 1** The schematic diagram of the experimental arrangement of LCOS SLM Phase Shifting Interferometer with Twyman- Green configuration with shifted mirror position to get blended fringes at the camera corners.

## 4. Four-step phase shift measurement

The relation between the applied gray value and the resulting phase shift from the measurements of the global phase shift was performed using a Matlab script based on the approach by Sun et al. [11] and the method described in our recently reported paper [14]. **Figure 2** shows the interferogram recorded at gray value 0 on the left-hand side and the surface reconstructed from a four-step PSA using the best gray values at 0, 60, 120 and 180 [13]. Zernike fitting and subtraction to isolate residual surface features in the measured optical phase data [12][13] is done and shown in **Figure 3**.

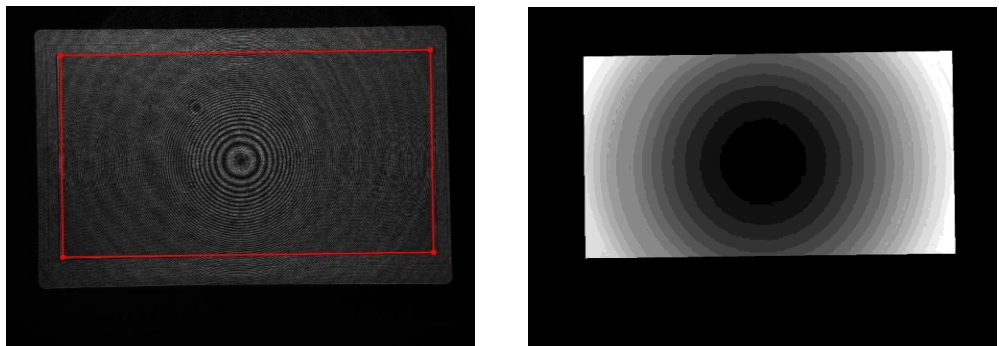

**Figure 2** Interferogram recorded at gray value 0 (left) and reconstructed surface from phase shift measurement (right).

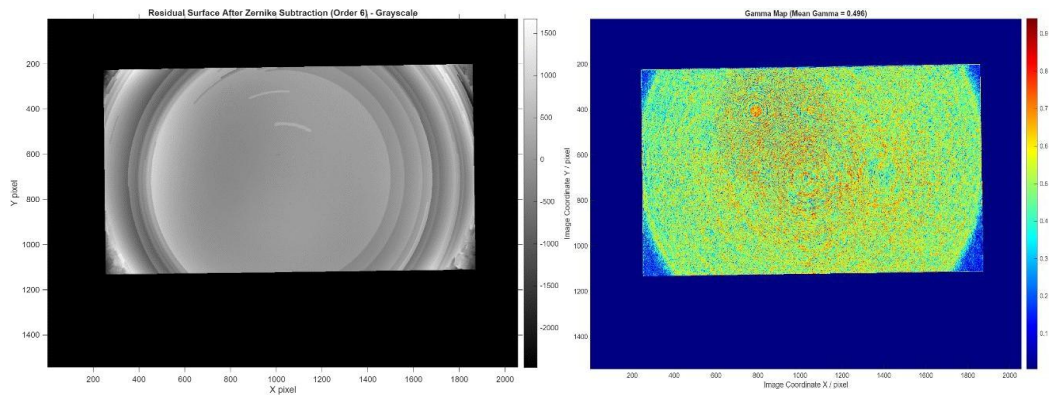

**Figure 3** Difference to a 6th order Zernike polynomial (left) and normalized fringe contrast 'Gamma' (right).

## 5. Wavefront Compensation

Wavefront compensation is done after a successful phase shift measurement and **Figure 4** shows the results obtained from the described workflow in our recent paper [12]. The left-hand side of **Figure 4** i.e., **(a)**, **(b)**, **(c)** and **(d)** show the interferograms recorded with a global phase shift of 0, 60, 120 and 180 with blended fringes at the camera corners and the right-hand side of *Figure 4* i.e., **(e)**, **(f)**, **(g)** and **(h)** show the interferograms with the inverse wrapped phase applied to the LCOS. As can be seen, due to the wavefront compensation, the intensity (and thus the phase) appears quite uniform within the active area of the LCOS. The thin lines resembling the interference pattern are the zones within the LCOS where a phase step between 0 and 2π occurs.

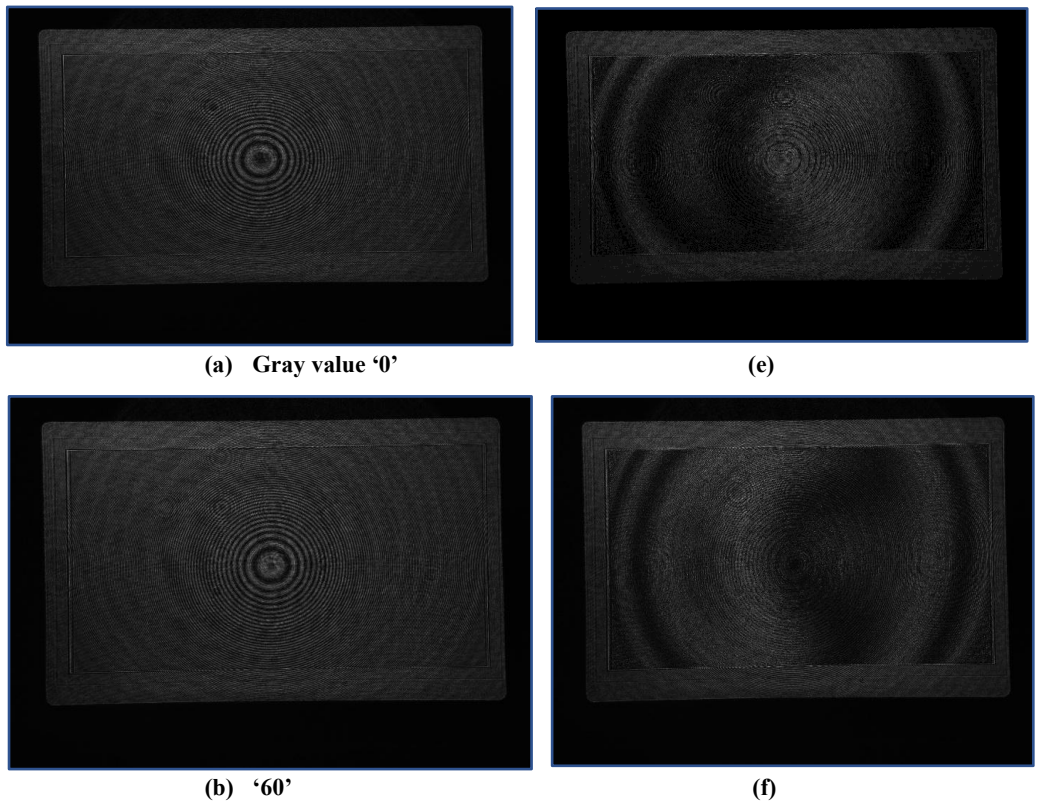

(a) Gray value '0'  (e)

(b) '60'  (f)

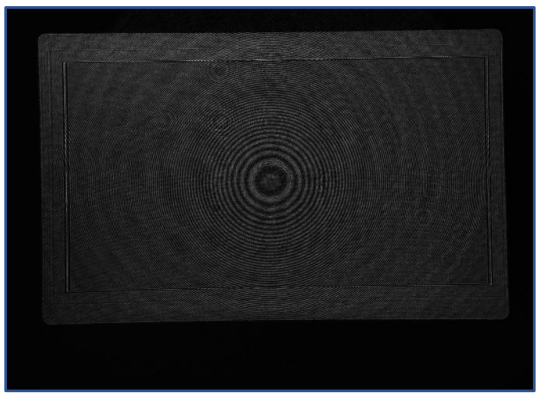
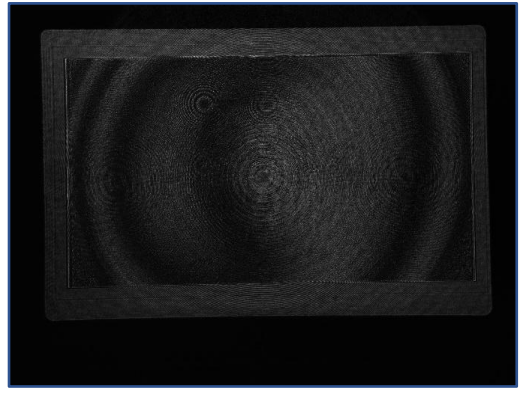

(c) '120'            (g)

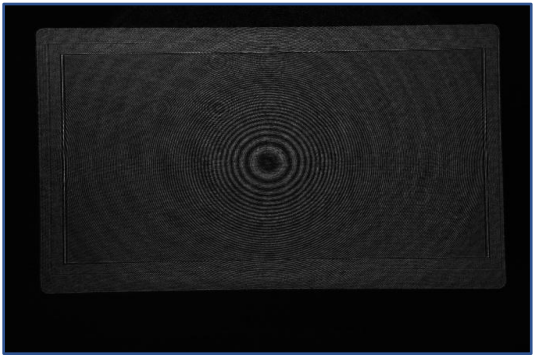
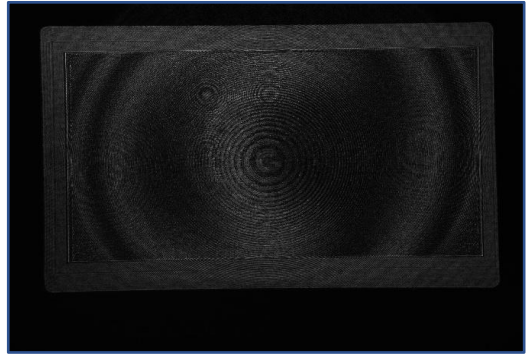

(d) '180'            (h)

**Figure 4**: **(a), (b), (c)** and **(d)** Interferograms are recorded at gray values 0, 60, 120, 180 without wavefront compensation (left) and **(e), (f), (g)** and **(h)** Interferograms recorded with (simple proof of concept) wavefront compensation determined from a four-step phase shift measurement (right).

The shape of the compensated wavefront is calculated using the four-step algorithm and it shows nearly a plane wave front. The 3D and 2D views of the reconstructed wavefront surfaces and their corresponding height maps before and after compensation are presented in **Figure 5 (a)**, **(b), (c), (d),** and **(f)** respectively.

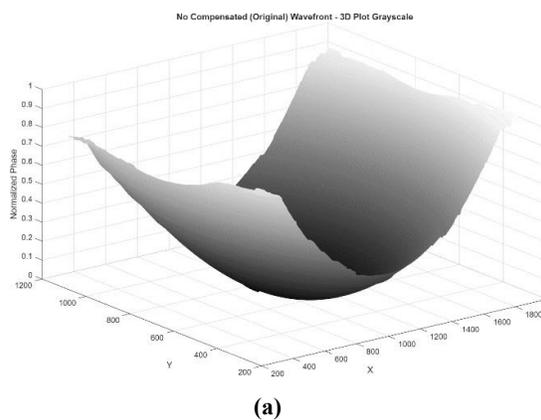
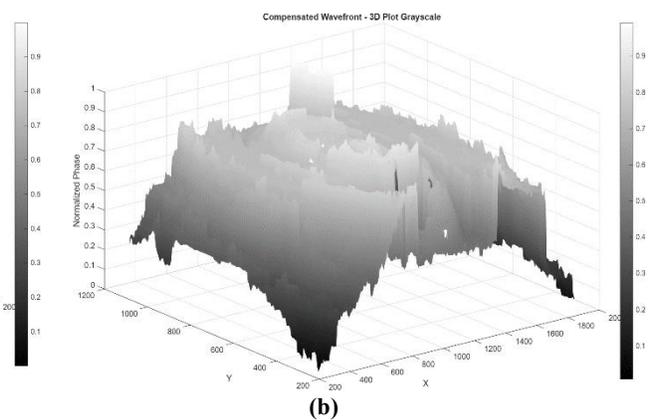

(a)            (b)

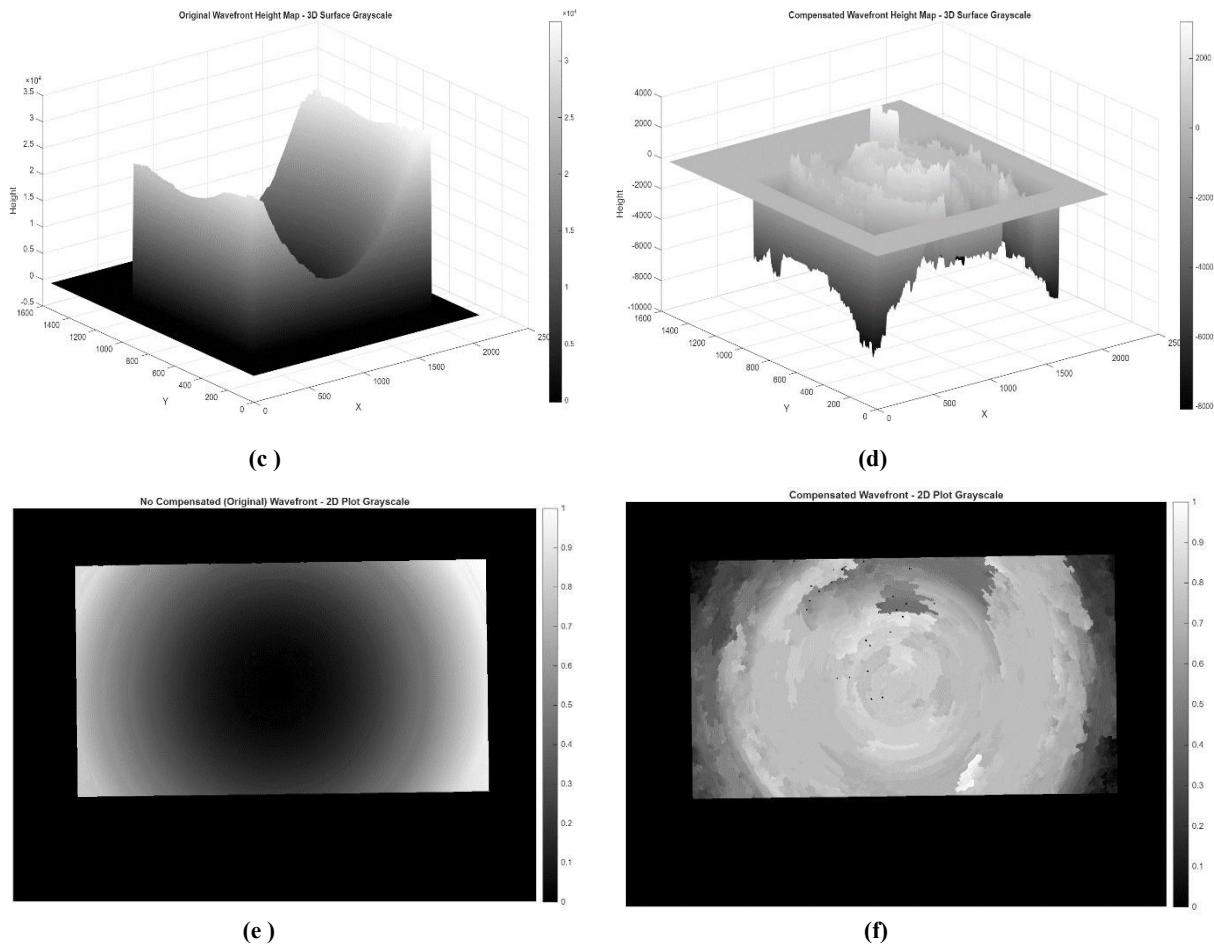

**Figure 5**: **(a)**, **(b), (c), (d), (e)and (f)**The 3D and 2D plots of the reconstructed wavefronts using four-step algorithm before and after compensation.

### 5. Conclusion and Discussions

Retrieval of the phase information using the 4-step Phase Shifting Algorithm (PSA) of a surface from an interferogram of high fringe density, especially in the corners, where the fringes may become unresolved, is suggested. The process is simple and it takes care of under-sampled, blended fringes.

### 6. Further scope of Research

Windowed Fourier Transform may be tried with adaptive window sizes to extract the phase locally in regions of high fringe density.

**Acknowledgement**

Indrani Bhattacharya[1] expresses sincere thanks and heartfelt gratitude to Prof. Rainer Tutsch for inviting and providing scope for an insightful and collaborative research in the field of 'LCOS SLM Interferometry with an application in Optical Metrology', in the Institute of Production Metrology (IPROM), *Techniche Universitat, Braunschweig, Germany*. She expresses sincere thanks and heartfelt gratitude to Dr. Marcus Petz for invaluable guidance and suggestions for pursuing this research work.

Sincere thanks and heartfelt gratitude to Ms. Annette Budin of IPROM to provide immense administrative help and guidance while pursuing the research work in the Institute.